\documentclass[10pt,a4paper,final]{IEEEtran}
\usepackage[english]{babel}
\usepackage[T1]{fontenc}
\usepackage{array}
\usepackage{graphicx}
\usepackage[caption=false,font=footnotesize]{subfig}
\usepackage{cite}
\usepackage{xfrac}
\usepackage{wrapfig} \usepackage{comment}
\usepackage[cmex10]{amsmath}
\usepackage{amsfonts,amssymb}
\usepackage{booktabs}
\usepackage{color}
\usepackage{epstopdf}
\usepackage{float}
\usepackage[T1]{fontenc}
\usepackage[cmex10]{amsmath}
\usepackage{amsfonts,amssymb}
\usepackage{multicol}
\usepackage{bm}
\usepackage{cite}
\usepackage{soul} 
\usepackage{url}
\usepackage[normalem]{ulem}
\usepackage[colorlinks=true, urlcolor=cyan, linkcolor=red, citecolor=red]{hyperref} 

\usepackage{fontawesome5}
\usepackage{colortbl} 
\usepackage{graphicx}
\usepackage{tabularray}
\usepackage{multicol}
\usepackage{multirow}
\usepackage{titlesec}
\titlespacing*{\subsection} {0pt}{1.7ex}{1.7ex}

\usepackage[linesnumbered,ruled,vlined]{algorithm2e}

\SetCommentSty{mycommfont}
\SetKwInput{KwInput}{Input}                
\SetKwInput{KwOutput}{Output}              

\usepackage{tikz}
\definecolor{gold}{rgb}{0.85,.66,0}

\begin{document}

\title{ Beamforming Control in RIS-Aided Wireless Communications: A Predictive Physics-Based Approach}

\author{{Luis C. Mathias}, 
{Atefeh Termehchi}, 
{Taufik Abr\~ao}, and
{Ekram Hossain}
\thanks{This work was partly supported by Brazil's National Council for Scientific and Technological Development (CNPq) under Grants 442945/2023-0 and 314618/2023-6.}
\thanks{L. C. Mathias is with the Computer Engineering Course (COENC). Federal Technological University of Paraná (UTFPR). Toledo, PR, Brazil, 85902-490. Email: \url{mathias@utfpr.edu.br}.
}
\thanks{T. Abr\~ao is with the Department of Electrical Engineering. The State University of Londrina (UEL). Po.Box 10.011; Londrina, PR, Brazil, 86057-970. Email: \url{taufik@uel.br}.
}
\thanks{Atefeh Termehchi and Ekram Hossain are with the Department of Electrical and Computer Engineering, University of Manitoba (UM), Winnipeg, MB, Canada, R3T 2N2. Email: \url{Atefeh.Termehchi@umanitoba.ca};  \url{Ekram.Hossain@umanitoba.ca}.}\\
\vspace{5mm}
}

\maketitle

\begin{abstract}
Integrating reconfigurable intelligent surfaces (RIS) into wireless communication systems is a promising approach for enhancing coverage and data rates by intelligently redirecting signals, through a process known as beamforming. However, the process of RIS beamforming (or passive beamforming) control is associated with multiple latency-inducing factors. As a result, by the time the beamforming is effectively updated, the channel conditions may have already changed. For example,
the low update rate of localization systems becomes a critical limitation, as a mobile UE’s position may change significantly between two consecutive measurements. To address this issue, this work proposes a practical and scalable physics-based solution that is effective across a wide range of UE movement models. Specifically, we propose a kinematic observer and predictor to enable proactive RIS control. From low-rate position estimates provided by a localizer, the kinematic observer infers the UE's speed and acceleration. These motion parameters are then used by a predictor to estimate the UE’s future positions at a higher rate, allowing the RIS to adjust promptly and compensate for inherent delays in both the RIS control and localization systems. Numerical results validate the effectiveness of the proposed approach, demonstrating real-time RIS adjustments with low computational complexity, even in scenarios involving rapid UE movement.
\end{abstract}


\section{Introduction}

Wireless communication systems supported by reconfigurable intelligent surfaces (RIS) can redirect signals to hard-to-reach areas, thereby improving coverage, reducing interference, and ultimately increasing both channel capacity and energy efficiency \cite{Hong2023}. However, as the number of RIS units, reflecting elements, base station (BS) antennas, and user equipment (UE) antennas increases, the overall system complexity is also expected to grow significantly \cite{He2022}.

A key challenge in deploying RIS-based systems is the need for real-time reconfiguration of the reflection coefficients of each RIS element as system conditions change. However, this task is complicated by the inherent latency in RIS setup procedures \cite{Saggese2023, Souza2024}. Several factors contribute to increased latency and reduced reconfiguration rates in RIS systems, including: the control communication method between the BS, RIS, and UE, whether through a dedicated control channel or by sharing the data channel \cite{Saggese2024}; the large number of RIS elements, which requires extensive computation for each adjustment; and delays introduced by digital-to-analog converters (DACs) and/or electronic multiplexers during the tuning of reverse voltages in the varactors of individual RIS elements \cite{Hong2023}. For instance, in the case of a mobile UE, these delays can lead to outdated beamforming configurations. By the time the controller updates the reflection coefficients, the UE may have already moved, making the adjustment ineffective. Even if the initial configuration is accurate, subsequent UE movement between updates can render it obsolete \cite{Panichpapiboon2009}. On the other hand, increasing the reconfiguration rate to keep up with user mobility can overload the control channel, reducing overall system throughput \cite{Saggese2024}. This issue is further exacerbated by localization system latency, which may cause even the initial RIS adjustment to be misaligned with the actual UE position.

Although global navigation satellite systems (GNSS), commonly found in smart devices, are widely used for positioning, they typically operate at sampling frequencies ranging from 1 to 20 Hz \cite{Gao2024,teseo2025}. These sampling frequencies are insufficient for real-time reconfiguration of RIS in many scenarios. 
Another approach to user localization is radio-based positioning using wireless networks, which is considered a fundamental capability in advanced 5G and 6G systems \cite{zheng2023jrcup}. Recent research highlights the potential of RIS-assisted localization in various scenarios, including those involving user mobility \cite{zhang2023active, zheng2023jrcup, teng2022bayesian, Hu2025}. However, these techniques often require estimation and search over a complex functional space whose dimensionality increases with the number of measurements \cite{teng2022bayesian}. This computational complexity poses a challenge for achieving real-time reconfiguration in RIS systems, hindering timely RIS adjustments, especially in scenarios requiring ultra-reliable low-latency communication (uRLLC). Furthermore, the beam-scanning RIS-assisted localization tend to have low acquisition rates \cite{wang2023b}.
This is because the beam-sweeping process occupies system resources, often preventing data transmission during the sweep. As a result, this approach involves a trade-off between the localization update rate and the average data transmission rate.

In addition, to address the challenge of efficient and timely RIS adjustments in the case of mobile UEs, \cite{Saggese2023} derives an upper bound on the outage probability by accounting for noise in UE position tracking within RIS-assisted URLLC systems. Based on this bound, the authors propose an energy- and computationally-efficient approach that enables control of the transmit power to meet URLLC reliability requirements. However, in the proposed approach, these requirements are satisfied by using higher-than-optimal transmit power to compensate for imperfections in the positioning information. 
 Another approach to address the challenge is to use prediction methods. Indeed, reliable prediction can compensate for the inherent delay in RIS control and enable real-time adjustment of the reflection coefficients. A very common solution for predicting the position of a mobile UE is to use deep learning (DL)-based approaches \cite{wang2023uav,hu2024location}. Authors in \cite{wang2023uav} propose a unmanned aerial vehicle (UAV) trajectory planning approach based on deep reinforcement learning (DRL) in combination with the prediction of UEs’ movement trend in future time slots. Indeed, extracting mobility features from the historical records of the UE and using a convolutional neural network allows the DRL model to adopt a future-oriented perspective. However, the prediction is limited to discrete movement directions (i.e. up, down, left, and right). Authors in \cite{hu2024location} propose a RIS-assisted localization prediction method that integrates Bayesian optimization with a long short-term memory (LSTM) network. However, as discussed earlier, the RIS-assisted localization approach has several drawbacks and may not be a practical solution for every scenario. Moreover, DL-based algorithms suffer from generalization limitations \cite{akrout2023domain}.   

In this context, we propose a predictive control approach for RIS beamforming to enable proactive RIS control. In particular, we propose RIS beamforming strategies based on observers and physics principles to mitigate performance degradation caused by RIS control delays and low-rate position measurements of mobile users.
We focus on an RIS-based wireless system, which includes a fixed RIS, a fixed base station (BS), and a mobile UE following an arbitrary and unknown trajectory.
Specifically, we introduce two complementary approaches: (i) the more complete observer-plus-predictor that considers speeds and accelerations (OPA), and (ii) its simplified variant, observer-plus-predictor that considers speeds but not accelerations (OPS).
These observers leverage a physics-based motion law to estimate the UE movement from sparse measurements, and to predict future positions at higher rates, thereby enabling proactive RIS control. We are able to estimate the UE position between two closely spaced time intervals because any nonlinear dynamics can be locally approximated as a linear one over sufficiently short durations \cite{khalil2015nonlinear}. Therefore, even if the UE follows a unknown nonlinear trajectory, the relatively short sampling interval allows for linearization, justifying the use of a physics-based motion model. In addition, unlike traditional Luenberger or Kalman observers that depend on a predefined trajectory model and output-error feedback, the proposed observers exploit physics-based motion laws to infer the UE’s movement and predict its future location. 

In summary, the contributions of this paper are as follows:
\begin{enumerate}
\item To address the low update rate of localization systems and the inherent delays in RIS control, we propose using a kinematic-based observer-plus-predictor. The kinematic state observer records the last measured position and estimates the speed and acceleration of the UE based on previous position measurements. Then, these speeds and accelerations are provided to the predictor, which predicts the UE’s position in advance and at higher rates. 
The proposed kinematic-based observer-plus-predictor is effective across a wide range of UE movements, even when the UE follows nonlinear and unknown trajectories.
Using the time-advanced position information, the RIS reflection coefficients are adjusted accordingly to ensure effective and timely beamforming.
\item The proposed approach is thoroughly evaluated across various UE movement scenarios, RIS control delays, sampling rates, and error levels of the secondary location system.
\item The proposed solution is of low complexity, easy to implement, and scalable, as it supports parallel position prediction for multiple UEs. 
For example, using time division, the RIS can dynamically reconfigure its beamforming pattern to sequentially serve multiple UEs. This enables the RIS to act as a multiplexer, allowing several users to access the BS in different time slots.
\end{enumerate}

The remainder of this paper is organized into five sections and an appendix.
The Section \ref{sec:ris_model} details the RIS modeling, the beamforming adjustment scheme, as well as the problem statement.
Section \ref{sec:kinematic} focuses on the modeling of the UE movement, as well as the design of the observer and predictor, incorporating the kinematics of the UE. Section \ref{sec:num_results} provides a numerical evaluation of the proposed predictive beamforming control approach. The key findings of this study and its conclusions are presented in Section \ref{sec:conclusion}. Finally, Appendix \ref{ap:ackscattering_block} provides details of the backscattering analysis used to determine the received power at the UE in the simulation section.

\section{System Model and Problem Statement}\label{sec:ris_model}
In this section, we first describe the considered BS-RIS-UE system model, then briefly present the electromagnetic model of the RIS required for the problem formulation, and finally define the problem. The problem focuses on designing a real-time RIS beamforming adjustment procedure that considers control latency and the low sampling rate of conventional UE localization methods.

\subsection{System Model and Assumptions} \label{subsec:system_model}

\begin{figure}[t]
    \centering
    \small  \includegraphics[width=.44\textwidth]{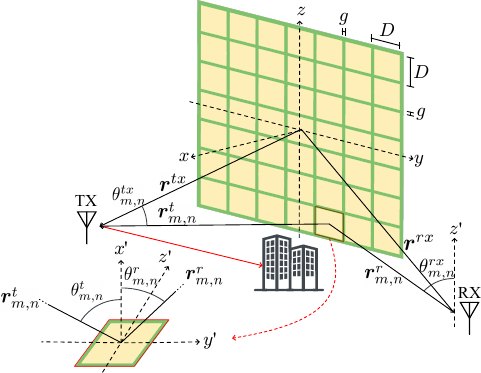} \vspace{-.25cm}
    \caption{ BS-RIS-UE model with notable vectors and angles.}
\label{fig:ris_model}
\end{figure}

As illustrated in Fig.~\ref{fig:ris_model}, we consider an RIS-aided wireless communication system consisting of a fixed BS, a fixed RIS, and a mobile UE. The RIS is modeled as an uniform planar array (UPA) with $M \times N$ elements, centered at the origin of the $y$-$z$ plane (${\bm r}^{\textsc{ris}} \!\!=\!\! [ 0\,\, 0\,\, 0]^{\textsc{t}}$m, where
$\{\cdot\} ^{\textsc{t}}$ denotes the transpose of a matrix or vector) and oriented along the $x$-axis  (${\bm u}^{\textsc{ris}} \!\!=\!\! [ 1\,\, 0\,\, 0]^{\textsc{t}}$). The position vectors of the BS and UE relative to the RIS center are denoted by $\bm{r}^{{tx}}$ and $\bm{r}^{{rx}}$, respectively. In addition, $\bm{r}^{t}_{m,n}$ and $\bm{r}^{{r}}_{m,n}$ denote the position vectors of the TX and RX regarding each ($m,n$)-th cell, respectively.

\begin{figure}[!t]
    \centering
    \footnotesize 
    \includegraphics[width=.42\textwidth]{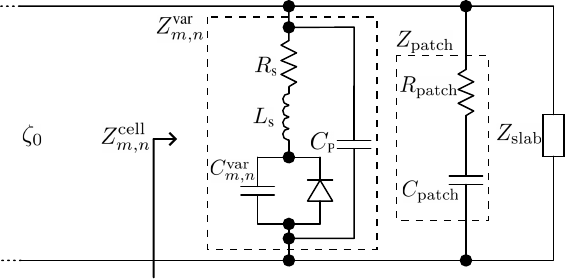}
    \\ (a) \vspace{.2cm} \\	
\includegraphics[width=.48\textwidth]{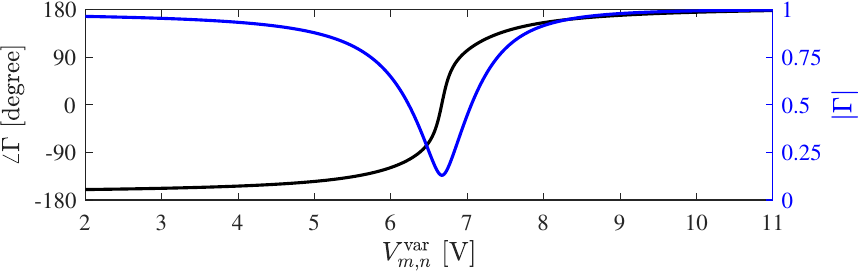}
    \\ (b) \vspace{-.2cm}\\ 
    \caption{Model of a RIS cell using transmission line analysis in (a) and magnitude and phase of the reflection coefficient magnitude of an RIS cell as a function of reverse DC voltage in (b).}
    \label{fig:lt_ris_model}
\end{figure}

As shown in Fig. \ref{fig:lt_ris_model}a, each RIS cell at row $m$, column $n$ can be modeled as a component of a transmission line \cite{Costa2021a,Costa2021b}. Therefore, each RIS element has its own reflection coefficient $\Gamma_{m,n}$, which depends on its input impedance $Z_{m,n}^{\text{cell}}$ via:
\begin{equation}
\Gamma_{m,n} = \frac{Z_{m,n}^{\text{cell}} - \zeta_0}{Z_{m,n}^{\text{cell}} + \zeta_0},
\label{eq:ref_coef}
\end{equation}
where $\zeta_0 = \sqrt{\mu_0 / \epsilon_0}$ is the free space impedance with the vacuum permittivity $\epsilon_0$ and magnetic permeability in vacuum $\mu_0$,  $Z_{m,n}^{\rm cell} = Z_{m,n}^{\text{var}} \parallel Z_{\rm patch} \parallel Z_{\rm slab}$ 
represents the RIS cell impedance, resulting from the parallel combination of the varactor, surface patch, and grounded dielectric impedances. $Z_{\rm patch}$ and $Z_{\rm slab}$ are constant for the RIS element considered,
and can be determined based on the relative permittivity $\epsilon_r$ and thickness $d$ of the uniform substrate surrounding the periodic structure, as well as the lattice periodicity $D$ and the gap $g$ between the squares along the $y$- and $z$-axes directions (Fig. \ref{fig:ris_model}) \cite{Costa2021a}.
The varactor impedance is determined by the reverse DC voltage $V_{m,n}^{\text{var}}$, which controls the junction capacitance \cite{skyworks2015}:
\begin{equation}
C_{m,n}^{\text{var}}(V_{m,n}^{\text{var}}) = 
C_{\textsc{jo}}\left(1+ \frac{V_{m,n}^{\text{var}}}{V_{\textsc{j}}}\right)^{-G},
\label{eq:jun_cap}
\end{equation}
where $C_{\textsc{jo}}$ denotes the zero-bias junction capacitance, $G$ is the grading coefficient, $V_{\textsc{j}}$ is the junction potential, and $V_{m,n}^{\rm var}$ is the reverse DC voltage. Indeed, $C_{\textsc{jo}}$, $V_{\textsc{j}}$, and $G$ are device-specific parameters.
By analyzing the circuit in Fig. \ref{fig:lt_ris_model}a, it is possible to determine the impedance of the varactor by $Z_{m,n}^{\text{var}} (C_{m,n}^{\rm var})=[R_{\text{s}} \parallel j\omega L_{\text{s}} \parallel (j\omega C_{m,n}^{\rm var})^{-1}]\parallel (j\omega C_{\text{p}})^{-1}$,
where $R_{\textsc{s}}$ and $L_{\textsc{s}}$ are the series resistance and inductance, and  $C_{\textsc{p}}$ is the package capacitance \cite{skyworks2015}.

The varactor capacitance, $C_{m,n}^{\text{var}}$, is the key tunable parameter in each RIS cell. By controlling it through the applied reverse DC voltage $V_{m,n}^{\text{var}}$, the RIS controller dynamically adjusts RIS cell’s impedance, $Z_{m,n}^{\rm cell} $, which in turn determines the reflection coefficient, $\Gamma_{m,n}$. This process is central to achieving the desired beamforming pattern and ensures that reflected signals coherently combine at the UE.
Therefore, with $k$ denoting the $k$-th time instant, the parameters of the RIS system can be arranged into $M \times N$ matrices and determined sequentially as follows:
\renewcommand{\theenumi}{\roman{enumi}}%
\begin{enumerate}
    \item varactor reverse DC voltages ${\bm V}^{\rm var}[k]$ are applied by the RIS controller, 
    \item varactor capacitances ${\bm C}^{\rm var}[k]$ are obtained by \eqref{eq:jun_cap},
    \item varactor impedances ${\bm Z}^{\rm var}[k]$ are obtained, 
    \item RIS cell impedances ${\bm Z}^{\rm cell}[k]$ are obtained,
    \item the coefficient reflection ${\bm \Gamma}[k]$ is specified by \eqref{eq:ref_coef}.
\end{enumerate}

It is worth noting that, to simplify this process, a look-up table (LUT) is used. This LUT stores precomputed values of $\Gamma$ as a function of $V^{\text{var}}$ based on the RIS model, as shown in Fig.~\ref{fig:lt_ris_model}b. During operation, the system can quickly retrieve the required voltage for the desired phase, reducing computational complexity. Fig.~\ref{fig:lt_ris_model}b was generated using the parameters listed in Table \ref{tab:param_cell}, based on \cite{Costa2021a} and the SMV1705 hyperabrupt junction varactor from Skyworks Inc. \cite{skyworks2015}, where $f$ and $\lambda$ denote the frequency and wavelength of the radio frequency (RF) signal.

\begin{table}[h!]
\centering
\caption{Parameters of one RIS cell and the SMV1705 varactor}
\vspace{-.2cm}
\label{tab:param_cell}
\begin{tabular}{|cc|cc|}
\hline
$f=3.5$~GHz             & $g=1$~mm      & $C_{\textsc{jo}}=31$~pF   & $R_{\textsc{s}} = 0.32 \, \Omega$ \\
$D=\frac{\lambda}{2}$   & $d = 1.57$~mm & $V_{\textsc{j}} =3$~V     & $L_{\textsc{s}}= 1.7$~nH \\
\multicolumn{2}{|c|}{$e_r=4.4-j0.088$}        & $G = 2$       & $C_{\textsc{p}}=0$~F   \\
\hline
\multicolumn{4}{|c|}{transverse magnetic (TM) mode with ${\bm r}^{tx} = [ 100 \, -\!100 \,\, 0]^{\textsc{t}}$~m} \\
\hline 
\end{tabular}
\end{table}

\subsection{Problem Statement} \label{sec:problem_statement}

To maximize received power at the UE, the reflected signals must combine constructively at the UE’s location. Assuming the UE position is available, the optimal reflection phase per RIS cell is computed by \cite{Costa2021a, Liu2021b}:
\begin{equation}
\angle \Gamma_{m,n}^{*} = \text{mod} \left( k_0 ( |\bm{r}^t_{m,n}| + |\bm{r}^r_{m,n}| ) + \pi, 2\pi \right) - \pi,
\label{eq:ang_ref_opt}
\end{equation}
where $k_0 = 2\pi / \lambda$ is the wave number and $\text{mod}(\varphi, 2\pi)$ denotes the modulo operation that wraps $\varphi$ into the interval $[0, 2\pi)$.

However, in practice, two main issues hinder real-time beamforming. The first is control latency, which arises from hardware and software delays. The second is the low sampling rate of UE localization systems, which prevents timely updates of the user's position. These delays result in outdated RIS settings that degrade signal quality. To overcome this, the goal of this paper is to precisely predict the UE’s future position at a higher rate than localization updates and adjust RIS beamforming proactively.
Therefore, in the next section, we propose a simple, physics-based solution that estimates the UE position at a high rate and predicts it ahead in time. Based on the predicted future position of the UE, the RIS cells are accordingly adjusted to achieve effective beamforming.

\section{Proposed Solution}\label{sec:kinematic}
In this section, we first model the UE's movement using a general nonlinear dynamic system. Then, we introduce the proposed physics-based observer and predictor for the UE's motion. Finally, we apply the observer and predictor that offers a straightforward, low-complexity solution for implementing predictive RIS beamforming.

\subsection{General Model of UE Movement} \label{subsec:mov_model}
 
We begin by modeling the UE's movement using a general discrete-time dynamic system:
\begin{align}
{\bf{x}}[k+1]&={\bf{f}}({\bf{x}}[k], {\bf{u}}[k]), 
\label{eq:gdyn}
\end{align}
where ${\bf{x}}[k]$ denotes state vector comprises the UE positions and corresponding speeds at time instant $k$, ${\bf{u}}[k]$ is the input vector at time instant $k$, ${\bf{f}}$ is an unknown, nonlinear (or linear), vector-valued function representing the UE’s motion dynamics.
Specifically, the state vector is defined as:
\begin{equation}
    {\bf{x}}[k] = \left[ {\begin{array}{*{20}{c}}
            {{p_{x,k}}}&{{p_{y,k}}}&{{p_{z,k}}}&v_{x,k}&v_{y,k}&v_{z,k}
    \end{array}} \right]^{\textsc{t}}.
\end{equation}

It is important to note that we consider the UE's movement as an autonomous dynamics, without any external control inputs. Nevertheless, 
\( \mathbf{u}[k] \) captures internal motion-driving factors, primarily acceleration, that influence the state transitions in accordance with physical laws.

If the UE moves linearly, its behavior can be modeled using basic kinematic equations\footnote{$p_{\cdot,k + 1} = p_{\cdot,k} + v_{\cdot,k}T + \frac{1}{2}a_{\cdot,k}T^2$ and $v_{\cdot,k + 1} = v_{\cdot,k} + a_{\cdot,k}T$.}. Therefore, assuming a  discretizations interval of $T$, the UE's motion dynamic can be represented as \cite{Becker2023book}:
\begin{align}
    {\bf{x}}[k + 1] &= {\bf{A}}{\bf{x}}[k] + {\bf{B}}{\bf{u}}[k],
    \label{eq:ss_linear}
\end{align}
which in expanded form results in:
\begin{equation*} \small
\begin{array}{l}
\left[ {\begin{array}{*{20}{l}}
{{p_{x,k + 1}}}\\
{{p_{y,k + 1}}}\\
{{p_{z,k + 1}}}\\
{{v_{x,k + 1}}}\\
{{v_{y,k + 1}}}\\
{{v_{z,k + 1}}}
\end{array}} \right] = \left[ {\begin{array}{*{20}{l}}
1&0&0&{T}&0&0\\
0&1&0&0&{T}&0\\
0&0&1&0&0&{T}\\
0&0&0&1&0&0\\
0&0&0&0&1&0\\
0&0&0&0&0&1
\end{array}} \right]\left[ {\begin{array}{*{20}{l}}
{p_{x,k}}\\
{p_{y,k}}\\
{p_{z,k}}\\
{{v_{x,k}}}\\
{{v_{y,k}}}\\
{{v_{z,k}}}
\end{array}} \right]\\
 \qquad \qquad \qquad \quad + \left[ {\begin{array}{*{20}{l}}
\frac{T^2}{2}&0&0\\
0&\frac{T^2}{2}&0\\
0&0&\frac{T^2}{2}\\
{T}&0&0\\
0&{T}&0\\
0&0&{T}
\end{array}} \right]\left[ {\begin{array}{*{20}{c}}
{{a_{x,k}}}\\
{{a_{y,k}}}\\
{{a_{z,k}}}
\end{array}} \right] .
\end{array}
\end{equation*}

In addition, the matrices in the linear dynamics can be rewritten in compact notation as:
\begin{equation}
    {{\bf{A}}} = \left[ \! {\begin{array}{*{20}{c}}
            {{{\bf I}_3}}&{{T\,{\bf{I}}_3}}\\
            {{{\bf{0}}_3}}&{{{\bf{I}}_3}}
    \end{array}} \! \right],\quad {{\bf{B}}} = \left[ \! {\begin{array}{*{20}{c}}
            \frac{T^2}{2} {{\bf{I}}_3}\\
            {{T\,{\bf{I}}_3}}
    \end{array}} \! \right],
\end{equation}
where ${\bf I}_\nu$ and ${\bf 0}_\nu$ are the identity matrix and the square matrix with all zero elements of dimension $\nu$, respectively.

\subsection{Physics-Based Observer and Predictor for UE Movement}

Nonlinear dynamics can be locally approximated as linear over sufficiently short time intervals~\cite{khalil2015nonlinear}. Thus, even if the actual motion of the UE follows nonlinear dynamics as described in \eqref{eq:gdyn}, a linear approximation can be adopted over short periods.
Consequently, a linear kinematic observer can be employed, allowing estimation of speeds and/or accelerations from previous UE position measurements, even if it follows an unknown non-linear trajectory.
These estimated variables are important for the operation of the subsequent stage, the predictor, which is intended to predict the UE position in advance in time.
Therefore, we design two kinematic observer-predictors: OPS, which estimates speed only, and OPA, which estimates both speed and acceleration. The input to both observers consists of noisy position measurements ${ \bf{y}}[q]$ sampled every $T_{\textsc{m}}$ seconds,
where $q=0,1,2,\ldots$ indexes denote the instants when the secondary localizer system provides a new position measurement. Accordingly, the input to the observer stage is defined as: 
\begin{equation}
    { \bf{y}}[q] = {{\bf{C}}}{\bf{x}}[q] +{\bf{n}}[q],
    \label{eq:measurement2_m}
\end{equation}
where ${{\bf{C}}} = \left[ \! {\begin{array}{*{20}{c}}
             {{{\bf{I}}_3}}&{{{\bf{0}}_3}}
    \end{array}} \! \right]$
is the measurement matrix and the measurement noise is typically modeled as ${\bf{n}}[q]\sim \mathcal{N}(0,{\bf R})$, with diagonal noise covariance matrix ${\bf R}= \text{diag}(\sigma_n^2,\,\sigma_n^2,\, \sigma_n^2)$ and noise variance $\sigma_n^2$. In addition, we define the correspondence between time indices $k$ and $q$ by:
\begin{equation}
  q=\lfloor J  k \rfloor,
  \label{eq:q}
\end{equation}
where $\lfloor.\rfloor$ is the floor function, and $J= \frac{T}{T_{\textsc{m}}}$ is the factor of the decimation process with $T_{\textsc{m}}\gg T$. 

Receiving position measurements every $T_{\textsc{m}}$ seconds may be short enough to linearize the UE dynamics. However, this may be a low rate for real-time RIS beamforming. 
To mitigate this problem, based on the speed and/or acceleration estimates made by the observer stage, the predictor stages increase the position sampling rate, making extrapolations of future positions in a period $T$ much shorter than $T_{\textsc{m}}$.

Having defined the observer’s input, we now explain the observer and predictor components of OPS and OPA. 
\subsubsection{Speed observer and predictor} The estimated speeds at instant  $q$ can be defined by $ {\bf{\hat v}}[q] = [\begin{matrix} {\hat v_{x,q}} & {\hat v_{y,q}} & {\hat v_{z,q}} \end{matrix}]^{\textsc{t}}$.
For $q=0$, there is no way to estimate the speed since there is no second measurement of the UE position yet.
Therefore, the observer stage initially considers the estimated speeds as zero, {i.e.}, $\bm{\hat v}[0]=[\begin{matrix}
0 & 0 & 0 \end{matrix}]^{\textsc{t}}$, in order to keep the position output of the predictor stage at a constant value from the first measurement to the subsequent measurement.
For $q \geq 1$, the observer stage estimates the speeds by:
\begin{equation}
{\bf{\hat v}}[q] = \frac{1}{T_{\textsc{m}}}
\left( {\bf{y}}[q] - {\bf{y}}[q-1] \right).
   \label{eq:OPS_v_2pt_m}
\end{equation}
At the predictor stage, the estimated UE position must be projected forward in time to account for the accumulated delay ${\hat T}_{\textsc{a}}$. This delay includes both the delay due to the low sampling rate of the localization system and the accumulated control delay, denoted by ${\hat t}_{\textsc{d}}$. The accumulated control delay refers to the time interval between the moment the UE position measurement is performed and the moment the corresponding electromagnetic adjustment of the RIS becomes effective. The estimate ${\hat t}_{\textsc{d}}$ consists of two components: (i) control delay and (ii) RIS setup delay. The control delay includes the time required for the UE position information to reach the controller, the processing time of this information, and the execution of the control algorithms. This component can typically be estimated by the control system designer. The RIS setup delay results from the timing mismatch between the DAC sampling rate and the demultiplexer circuit responsible for distributing a large number of reverse DC voltages to the varactors. It also accounts for the stabilization time of these voltages. This delay is generally characterized through empirical tests conducted by the RIS manufacturer.

Accordingly, the last measurement, the estimated speed and the estimated accumulated delay are used as parameters that allow predicting the UE positions at a higher sampling rate and advanced in time by the OPS predictor equation:
\begin{equation}
   {\hat {\bm r}}^{rx}[k] = {\bf{y}}[q] + {\bf{\hat v}}[q] {\hat T}_{\textsc{a}}[k],
   \label{eq:ops_pred_m}
\end{equation}
with an estimated time in advance of:
\begin{equation}
{\hat T}_{\textsc{a}}[k]  =  \left( k-\lfloor J^{-1} q \rfloor \right)T +{\hat t}_{\textsc{d}}.
    \label{eq:ta}
\end{equation}

\subsubsection{Speed and acceleration observer and predictor} 
The OPA observer also operates in a staggered manner, similar to the OPS observer. It initializes the speed values as zero for $q=0$,  and estimates them using \eqref{eq:OPS_v_2pt_m} for $q=1$. 
However, \eqref{eq:OPS_v_2pt_m} presents the average speed between the two position measurements. 
Thus, after the third measurement ($q \geq 2$), a three-point backward-difference expression is used for the last three positions measurement and returns the instantaneous speed at the $q$ instant \cite{Wu1990}:
\begin{equation}
   {\bf{\hat v}}[q] = \frac{1}{2T_{\textsc{m}}}
\left(
   3 {\bf{y}}[q]
   -4 {\bf{y}}[q-1]
   + {\bf{y}}[q-2]
\right).
   \label{eq:opa_v_3pt_m}
\end{equation}
Another difference in the OPA observer is that the accelerations $ {\bf{\hat a}}[q] = [\begin{matrix} {\hat a_{x,q}} & {\hat a_{y,q}} & {\hat a_{z,q}} \end{matrix}]^{\textsc{t}}$ are also considered.
As long as $q<2$, there is no third localizer measurement, and therefore the accelerations cannot be estimated. In this case they are also considered zero, \textit{i.e.} $ {\bf{\hat a}}[0]={\bf{\hat a}}[1] = [\begin{matrix} 0 & 0 & 0 \end{matrix}]^{\textsc{t}}$.
After that, in addition to obtaining the speeds, the OPA observer obtains the acceleration for $q \geq 2$ by \cite{Carneiro2016}:
\begin{equation}
   {\bf{\hat a}}[q] = \frac{1}{T_{\textsc{m}}^2}
\left(
   {\bf{y}}[q]
   -2 {\bf{y}}[q-1]
   + {\bf{y}}[q-2]
\right),
   \label{eq:opa_accel_m}
\end{equation}
allowing the construction of a more complete predictor by:
\begin{equation}
   {\hat {\bm r}}^{rx}[k] = {\bf{y}}[q] + {\bf{\hat v}}[q]{\hat T}_{\textsc{a}}[k] + \frac{1}{2}{\bf{\hat a}}[q]{\hat T}_{\textsc{a}}[k]^2 .
   \label{eq:opa_pred_m}
\end{equation}
In this framework, the OPA predictor assumes constant acceleration, whereas the OPS predictor assumes constant speed between two consecutive observer updates. This assumption is used to predict the UE's position ${\hat {\bm r}}^{rx}[k]$ at a higher temporal resolution. As a result, the RIS can perform beamforming adjustments in a more timely and accurate manner.
The pseudocode is presented in \textbf{Algorithm \ref{alg:OPS_opa}}, where the variable $\textit{mode}$ determines whether the solution operates under the OPS or OPA approach. Note that lines 23 and 24 of the pseudocode include the RIS beamforming adjustment step, which will be described in detail in the following subsection.

\newcommand{\algrule}[1][.2pt]{\par\vskip.5\baselineskip\hrule height #1\par\vskip.5\baselineskip}
\begin{algorithm}[h] \footnotesize
\DontPrintSemicolon 
\KwData{${mode}$, $T_{\textsc{m}}$, $T$, ${\hat t}_{\textsc{d}}$, $k_0$, $\bm{r}^{{tx}}$, \textit{3-D position measurement of UE} }
\KwResult{${\bm V}^{\rm var}$}
\algrule
${\bf{y}}_a \leftarrow [0\,0\,0]^{\textsc{t}}$;
${\bf{y}}_b \leftarrow [0\,0\,0]^{\textsc{t}}$;
${\bf{y}}_c \leftarrow [0\,0\,0]^{\textsc{t}}$\\
${\bf{\hat v}} \leftarrow [0\,0\,0]^{\textsc{t}}$ \\
${\bf{\hat a}} \leftarrow [0\,0\,0]^{\textsc{t}}$ \\
$J \leftarrow \frac{T} {T_{\textsc{m}}}$\\
$k \leftarrow 0$ \\
$q \leftarrow -1$ \\
\While{true}
{
    \tcp{At each period $T$ executes}    
    \If{received new measurement of the UE position}
    {   
        \tcp{At each period $T_{\textsc{m}}$ executes}
        $q \leftarrow q+1$ \\
        \tcp{Observer stage}
        ${\bf{y}}_c \leftarrow {\bf{y}}_b $ \\
        ${\bf{y}}_b \leftarrow {\bf{y}}_a $ \\
        ${\bf{y}}_a \leftarrow $ \textit{3-D position measurement of UE} \\
        \If{$(q=1) \,||\, \left(\,({mode}={\rm OPS}) \,\&\&\, (q > 1)\,\right)$}
           {
               ${\bf{\hat v}} \leftarrow \frac{1}{T_{\textsc{m}}} \left( {\bf{y}}_a - {\bf{y}}_b \right)$ \tcp*{\eqref{eq:OPS_v_2pt_m}}
           }
        \Else 
        { 
            ${\bf{\hat v}} \leftarrow \frac{1}{2T_{\textsc{m}}} \left( 3 {\bf{y}}_a -4 {\bf{y}}_b + {\bf{y}}_c \right)$ \tcp*       {\eqref{eq:opa_v_3pt_m}}
            ${\bf{\hat a}} \leftarrow \frac{1}{T_{\textsc{m}}^2} \left({\bf{y}}_a -2 {\bf{y}}_b + {\bf{y}}_c \right)$ \tcp*{\eqref{eq:opa_accel_m}}
        }
    }
    \tcp{Predictor stage}
    ${\hat T}_{\textsc{a}}  \leftarrow  \left( k-\lfloor J^{-1} q \rfloor \right)T +{\hat t}_{\textsc{d}}$ \tcp*{\eqref{eq:ta}} 
    \If{${mode}={\rm OPS}$}
    {
        ${\hat {\bm r}}^{rx} \leftarrow {\bf{y}}_a + {\bf{\hat v}} {\hat T}_{\textsc{a}}$ \tcp*{\eqref{eq:ops_pred_m}}
    }
    \Else
    {
        ${\hat {\bm r}}^{rx} \leftarrow {\bf{y}}_a + {\bf{\hat v}}{\hat T}_{\textsc{a}} + \frac{1}{2}{\bf{\hat a}}{\hat T}_{\textsc{a}}^2$ \tcp*{\eqref{eq:opa_pred_m}}
    }
    \tcp{RIS beamforming}
    Estimates $\angle{\bm \Gamma}$ with $k_0$, $\bm{r}^{{tx}}$ and ${\bm {\hat r}}^{rx}$. \tcp*{\eqref{eq:ang_ref_opt}}
    ${\bm V}^{\rm var} \leftarrow \text{LUT} \left( \angle{\bm \Gamma} \right)$ \\
    \vspace{.1cm}
    $k \leftarrow k + 1$ \\
    Wait for period $T$. \\
}
\caption{\small OPS and OPA solution}
\label{alg:OPS_opa}
\end{algorithm}

\subsection{Proposed approach for Predictive Real-time RIS Beamforming}

\begin{figure*}[t]
\centering
\footnotesize
\includegraphics[width=.98\textwidth] {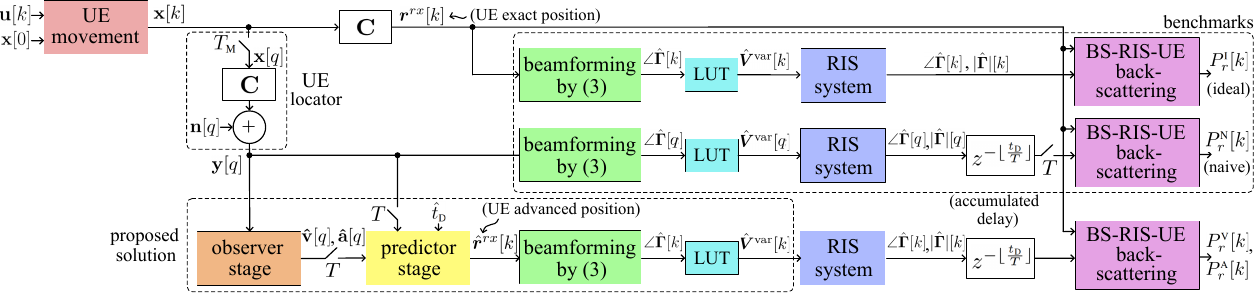}
\caption{
Arrangement used to evaluate the received power at the UE. The ideal benchmark assumes perfect knowledge of the UE position and ignores any delay introduced by the RIS control system. In contrast, both the naive and the proposed approaches consider realistic conditions, including noisy position measurements and the presence of accumulated control delays. The proposed observer-plus-predictor approach compensates for both the control delay and the low update rate of the secondary locator system.
}
\label{fig:ris_proposed_solution}
\end{figure*}

In this subsection, we explain how the UE positions predicted by the OPS and OPA approaches enable proactive RIS control.
Considering the BS and RIS in known fixed positions, the implementation method of the proposed solution is outlined in Fig. \ref{fig:ris_proposed_solution}.  This figure also explains how the benchmarks that will serve as a performance comparison in terms of power captured in the UE are obtained.

In the proposed approach, the kinematic model emulates the UE movement ${\bf{x}}[k]$ from the initial state vector ${\bf{x}}[0]$ and the input acceleration vector ${\bf{u}}[k]$.
This signal is resampled at a lower rate (sampling period $T_{\textsc{m}}$) and multiplied by the measurement matrix ${\bf{C}}$ to extract the exact positions of the UE in the vector ${\bf{x}}[q]$.
Next, the noise ${\bf{n}}[q]$ is added, generating the signal ${\bf{y}}[q]$.
After, the UE observer stage estimates its speeds and/or accelerations at each period $T_{\textsc{m}}$. 
Based on these parameters, the predictor estimates the UE position $\hat {\bm r}^{rx}[k]$, explicitly incorporating the accumulated delay ${\hat t}_{\textsc{d}}$.
Then, $\hat {\bm r}^{rx}[k]$ is used in the beamforming design according to \eqref{eq:ang_ref_opt}, which determines the reflection phase of each RIS element.
These reflectance phases allow determining the reverse DC voltages of the varactors using the LUT presented at the end of Subsection \ref{subsec:system_model}, allowing to control 
RIS system block.

\subsection{Computational Complexity, Scalability, and Implementation}

After initializing the observers ($q \geq 2$), Table \ref{tab:complexity} summarizes the computational complexities of each stage of the proposed approach, expressed as the number of multiplications or divisions per second. It can be observed that the computational burden of the observer and predictor stages is negligible compared to the significantly higher cost of computing the phase shifts of each RIS element via \eqref{eq:ang_ref_opt} and subsequently converting these values into voltages for the corresponding varactors through LUT access. The low complexity of the observer stages makes the proposed approach scalable to multiple UEs connected to the BS, as the controller can estimate the motion parameters of several UEs in parallel. By scheduling the UEs in advance, the BS can enable multiplexing through dynamic switching of RIS beamforming, thereby realizing time-division multiple access (TDMA). In this framework, the predictor and subsequent stages operate only on the movement parameters of the UE scheduled for the current time slot, enabling the predictive beamforming adjustment proposed in this work.

\renewcommand{\arraystretch}{1.4}
\begin{table}[]
\centering
\caption{Complexity in number of multiplications or divisions}
\vspace{-.2cm}
\label{tab:complexity}
\begin{tabular}{|c|cc|cc|}
\hline
\multirow{2}{*}{Stage} & \multicolumn{2}{c|}{General}                                                 & \multicolumn{2}{c|}{Simulated$^*$} 
\\ [.1 em] 
                       & \multicolumn{1}{c|}{OPS}                        & OPA                        & \multicolumn{1}{c|}{OPS}   & OPA   
                       \\ [.2 em] \hline
Observer               & \multicolumn{1}{c|}{$\frac{3}{T_{\textsc{m}}}$} & $\frac{6}{T_{\textsc{m}}}$ & \multicolumn{1}{c|}{$30$}    & $60$    
\\ [.2 em] \hline
Predictor              & \multicolumn{1}{c|}{$\frac{5}{T}$}              & $\frac{10}{T}$             & \multicolumn{1}{c|}{$5\times10^{3}$}  & $10\times10^{3}$ 
\\ [.1 em] \hline
\eqref{eq:ang_ref_opt} & \multicolumn{2}{c|}{$2\frac{MN}{T}$}                                          & \multicolumn{2}{c|}{$1.8\times10^{6}$}   
\\ [.1 em] \hline
LUT                    & \multicolumn{2}{c|}{$\frac{MN}{T}$}                                          & \multicolumn{2}{c|}{$0.9\times10^{6}$}    
\\ [.05 em] \hline
\end{tabular}
\\ \vspace{.2 em} {\footnotesize $^*$For $T_{\textsc{m}}\!=\!100$~ms, $T\!=\!1$~ms and $M\!=\!N\!=30$.}
\end{table}

Another relevant aspect is the progressive increase in the amount of data processed at each stage. In the first stage, the observer operates with a small set of parameters (position, velocity, and/or acceleration) and at a low update rate. In contrast, subsequent stages operate at higher rates and with larger data structures (matrices $M \times N$), which significantly increases the volume of data processed per second. To alleviate the data transmission load on the control channel, it is advantageous to execute the algorithms in the final stages within a controller colocated with the RIS. In this configuration, only UE location measurements and scheduling information are exchanged between the BS, RIS, and UEs, while the more complex and memory-intensive computations are performed locally at the RIS. This combination of low observer complexity and localized processing ensures scalability, enabling the RIS to dynamically and proactively adapt its resources to serve multiple UEs simply and efficiently.

\section{Simulations and Numerical Results} \label{sec:num_results}

This section evaluates the proposed RIS predictive control approach, which aims to maximize the received power of a mobile UE via RIS beamforming.

\subsection{Simulation Parameters, Settings, and Benchmarking}

The arrangement used to evaluate the received power at the UE is illustrated in Fig.~\ref{fig:ris_proposed_solution}. As shown in the figure, the magnitudes and phases of each RIS element are first applied to a block that represents the accumulated delay in RIS beamforming, and subsequently to the BS–RIS–UE backscattering block defined in Appendix~\ref{ap:ackscattering_block}. This block emulates the received power $P_r[k]$ at the exact position of the UE, given by ${\bm r}^{rx}[k] = {\bf C}{\bf x}[k],
$ but with the RIS beamforming adjusted according to \eqref{eq:ang_ref_opt} using the estimated UE position ${\hat{\bm r}}^{rx}[k]$.

Note that the proposed simple and complete solution, with very low complexity, allows to predict the UE position in the future with a higher rate, compensating the $t_{\textsc{d}}$ delay and allowing the proper adjustment of the RIS in real time.
The received power at the UE is evaluated for the two proposed observer-plus-predictor approaches: $P_r^{\textsc{s}}[k]$ for the OPS and $P_r^{\textsc{a}}[k]$ for the OPA.
To evaluate performance, the obtained results are compared against two benchmarks.
The first considers the power in the ideal case $P_r^{\textsc{i}}[k]$ in which the RIS is instantaneously adjusted to the exact UE position in ${\bm r}^{rx}[k]$ and without delays, which is not possible in practice.
The second benchmark implements a naive RIS control, applied only after each low-rate sample affected by  noise from the locator system and with beamforming delayed by $t_{\textsc{d}}$ seconds. The received power resulting from this strategy is denoted as $P_r^{\textsc{n}}[k]$.
All powers obtained in this work are in dBm scale.

\begin{figure}[t]
\centering
\small 
\includegraphics[width=.46\textwidth]{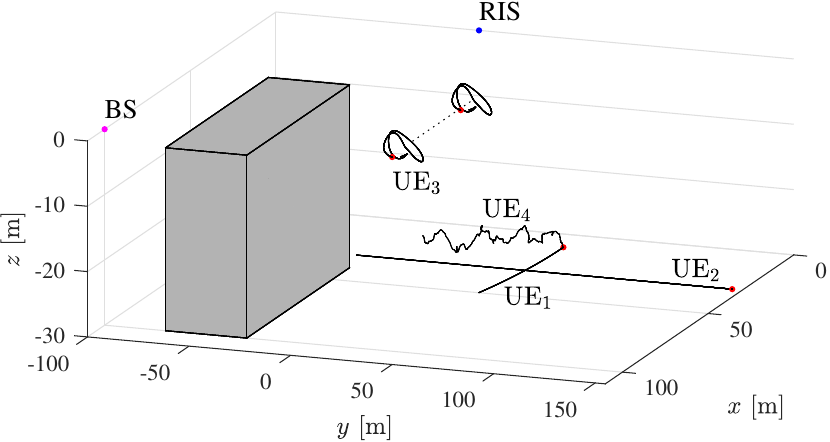}
\vspace{-.2cm}
\caption{BS-RIS-UE scenarios with different UE movement patterns: UE$_1$ (constant acceleration), UE$_2$ (real car acceleration data), UE$_3$ (drone trajectory under aggressive piloting), and UE$_4$ (Gauss-Markov mobility model). The red dot indicates the UE's starting position, and the black line shows its trajectory.}
\label{fig:ris_scenario}
\end{figure}

For the numerical simulations, in addition to the RIS cell parameters listed in Table~\ref{tab:param_cell}, Table~\ref{tab:param} provides the general system parameters. Moreover, Fig.~\ref{fig:ris_scenario} illustrates the positions of the fixed BS, fixed RIS, and mobile UE across four distinct scenarios, each following a unique trajectory:
\begin{itemize}
\item[UE$_1$] The simplest case, moving with constant acceleration along the $y$-axis and constant speed along the $x$-axis, serving as a clear baseline to validate the proposed control approach;
\item[UE$_2$] Follows a straight path, incorporating real acceleration data from 0 to 100 km/h tests of both combustion and electric cars;
\item[UE$_3$] Represents the most complex scenario, tracking a drone’s trajectory during an aggressive race and evaluated at varying distances from the RIS along the $x$-axis;
\item[UE$_4$] The Gauss-Markov mobility model is considered, which allows evaluation under different levels of memory and randomness in the movement.
\end{itemize}

\begin{table}[h!]
\centering
\caption{General parameters of numerical simulations} 
\vspace{-.2cm}
\label{tab:param}
\begin{tblr}{
  hline{1} = {-}{},
  hline{5} = {-}{},
  hline{4} = {2}{},
  hline{4} = {3}{},
  vline{1} = {-}{},
  vline{2} = {-}{},
  vline{3} = {1-3}{},
  vline{4} = {-}{},
}
general: & BS: & RIS:  \\
$T=1$~ms                    & $P_t=50$~W    & $M=30$ \\
$T_{\textsc{m}}=100$~ms     & $q=100$       & $N=30$ \\
$t_{\textsc{d}}=20$~ms               & {{\SetCell[2]{l} UE: half-wave dipole antenna}}         
\end{tblr}
\end{table}

\subsection{Numerical Results: Constant Speed and Acceleration Scenarios}

The first scenario considers an initial UE$_1$ position of $[\begin{matrix}
10 & 50 & -30 \end{matrix}]^{\textsc{t}}$~m, non-zero initial speed only on the $x$-axis of $v_x=20$~m/s, \textit{i.e.} ${\bf{x}}[0]=[\begin{matrix} 10 & 50 & -30 & 20 & 0 & 0\end{matrix}]^{\textsc{t}}$, and constant acceleration only in $y$-axis, {i.e.}, ${\bf{u}} =[\begin{matrix}
0 & -4 & 0 \end{matrix}]^{\textsc{t}}$~m/s$^2$.
The speed and acceleration parameters are typical values in current automobiles \cite{Meyn2015}.
These parameters emulate a UE$_1$ motion on the $x$-$y$ plane ($v_z=0$~m/s), with constant speed on the $x$-axis and uniformly accelerated on the $y$-axis.
In the observer stages of OPS and OPA, a position measurement rate of 10 samples per second (Sa/s) was considered, implying a period of $T_{\textsc{m}}=100$~ms, while the model and predictor stage used a rate of 1~kSa/s ($T=1$~ms).

\begin{figure}[t]
    \centering
    \footnotesize
    \includegraphics[width=.48\textwidth]{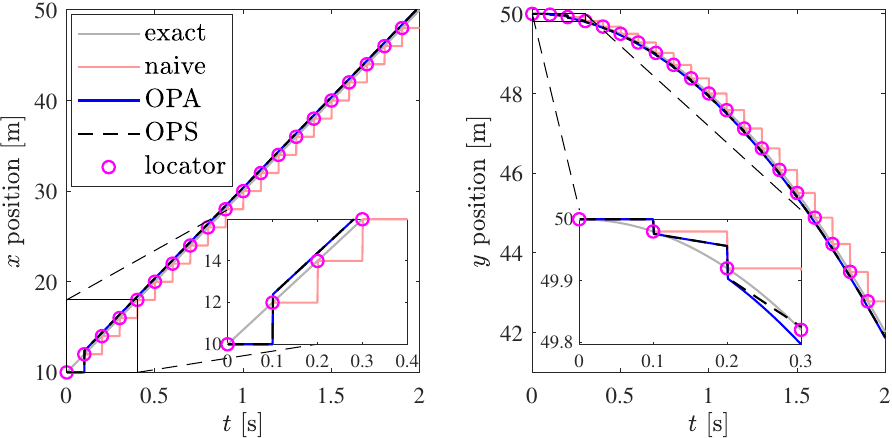}
    \\ \vspace{-.25cm} (a) \vspace{.25cm} \\	\includegraphics[width=.48\textwidth]{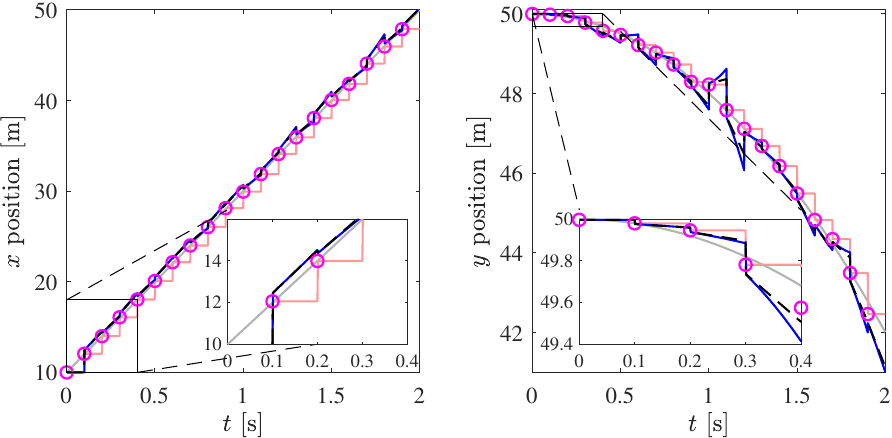}
    \\ \vspace{-.25cm} (b) \vspace{.25cm} \\		\includegraphics[width=.44\textwidth]{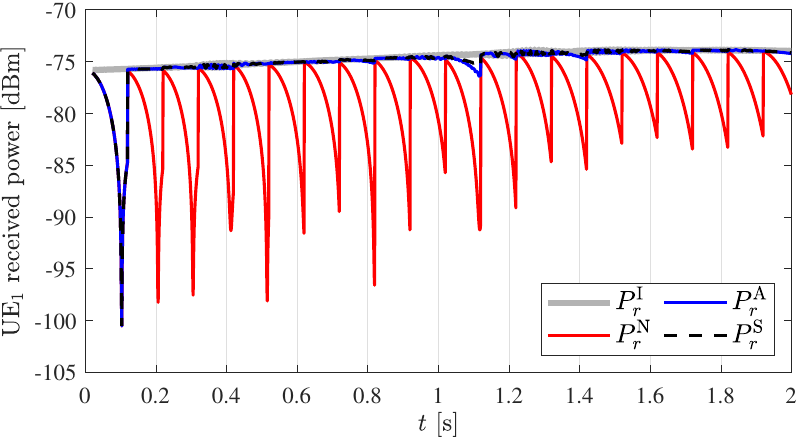}
    \\ (c) \vspace{-.2cm} \\
    \caption{
    Trajectory of UE$_1$ in the $x$ and $y$-coordinates obtained using the exact model, the naive approach, and the observer-plus-predictor approaches: one that accounts for both speed and acceleration (OPA), and another that considers only speed (OPS).
    In (a) ideal condition without noise and in (b) with noise in the position measurement with $\sigma_n=0.1$~m.
    In (c), power captured in UE$_1$ for the different approaches considering trajectories in (b).}
    \label{fig:Pr}
\end{figure}


\subsubsection{Accuracy of prediction}
For the $x$ and $y$ coordinates, Figs.~\ref{fig:Pr}a and \ref{fig:Pr}b illustrate the exact trajectory, the low-rate measured positions, and the naive and high-rate predicted trajectories obtained with the OPS and OPA approaches. 
Fig.~\ref{fig:Pr}a presents an ideal scenario with no measurement errors, while Fig.~\ref{fig:Pr}b considers a measurement error with standard deviation $\sigma_n = 0.1$~m. These figures show that the proposed OPS and OPA predictors estimate the UE$_1$ trajectory ahead in time. 
This capability allows them to mitigate the accumulated delay in RIS beamforming control. In Fig.~\ref{fig:Pr}a, the performance of the predictors, including the naive approach, is identical between the first and second position measurements of the localizer. 
Between the second and third measurements, the OPA and OPS estimators yield equally smaller errors compared to the naive predictor. 
From the third measurement onwards, for the movement with constant speed along the $x$-axis, the performance of OPA and OPS remains similar. 
However, for the movement with non-zero acceleration along the $y$-axis, the OPA achieves smaller errors since it explicitly accounts for acceleration in both its observer and predictor stages. 

\subsubsection{Predictive control of beamforming}
From the trajectories obtained with position measurement noise in Fig.~\ref{fig:Pr}b, Fig.~\ref{fig:Pr}c shows the power captured by UE$_1$ in dBm. 
The curves correspond to the ideal instantaneous adjustment of the RIS ($P_r^{\textsc{i}}$), the naive adjustment performed with delay $t_{\textsc{d}}$ and only at each localizer sampling ($P_r^{\textsc{n}}$), and the proposed approaches that combine observers with predictors ($P_r^{\textsc{a}}$ and $P_r^{\textsc{s}}$). When compared to the ideal curve $P_r^{\textsc{i}}$, the naive case ($P_r^{\textsc{n}}$) deteriorates significantly over time after delayed adjustments, as UE$_1$ moves. 
For example, at $t \simeq 0.7$~s, its performance is about 24~dB worse than the ideal received power. 
Such deep power valleys make it unfeasible to satisfy uRLLC requirements. By contrast, the observer-plus-predictor solutions achieve performance close to the ideal curve after only two samples, corroborating the effectiveness of the proposed approach.  
Although the degradation in position estimates due to localizer noise is evident in Fig.~\ref{fig:Pr}b, the received power performance of all proposed approaches in Fig.~\ref{fig:Pr}c is not significantly affected, demonstrating the robustness of the system.
\subsubsection{Noise level analysis}
\begin{figure}[t]
    \centering
    \small \includegraphics[width=.44\textwidth]{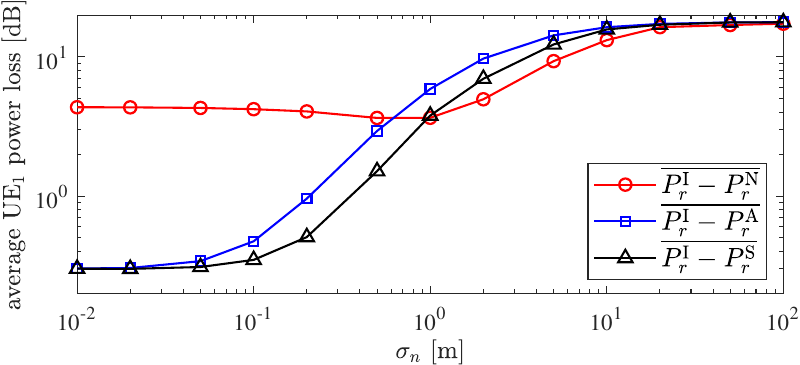}
    \vspace{-.25cm}
    \caption{UE$_1$ average power loss of the three control schemes in relation to the ideal control, all as a function of the standard deviation of the localizer position measurement.
    }
    \label{fig:PLs}
\end{figure}

Next, Fig.~\ref{fig:PLs} shows the average power loss as a function of the noise standard deviation in the location, considering the same scenario as in Fig.~\ref{fig:Pr}. 
For each evaluated point, 100 realizations are used. It can be observed that the proposed OPS and OPA approaches incur losses close to 0.3~dB for $\sigma_n < 0.2$~m, while the naive scheme exhibits an average loss slightly higher than 4.3~dB. 
For $\sigma_n > 0.2$~m, the performance of OPS and OPA begins to degrade, with OPS outperforming OPA. 
This highlights the greater robustness of the simpler OPS scheme under higher levels of location measurement noise\footnote{Considering $T_{\textsc{m}} < 1$, the acceleration estimate is more sensitive to noise in the position measurement due to the factor $T_{\textsc{m}}^{-2}$ in \eqref{eq:opa_accel_m}, whereas for the velocity estimate in OPS the factor is much smaller, $T_{\textsc{m}}^{-1}$ in \eqref{eq:OPS_v_2pt_m}.}. The OPS approach outperforms the naive benchmark up to $\sigma_n = 1$~m. 
Beyond this point, both predictor approaches experience significant performance degradation, as larger noise in the location measurements propagates into higher errors in the speed and acceleration estimates, thereby reducing the accuracy of the future position predicted. 
For $\sigma_n > 20$~m, all approaches converge to an average loss level of 17.8~dB.

These numerical results validate the proposed solution for applications that employ precise point positioning (PPP) localization, which provides centimeter-level accuracy. 
PPP is increasingly adopted in autonomous vehicle navigation and other precision-demanding fields~\cite{An2023,teseo2025}.

\subsubsection{Different locator sampling periods}
\begin{figure}[t]
    \centering
    \footnotesize \includegraphics[width=.44\textwidth]{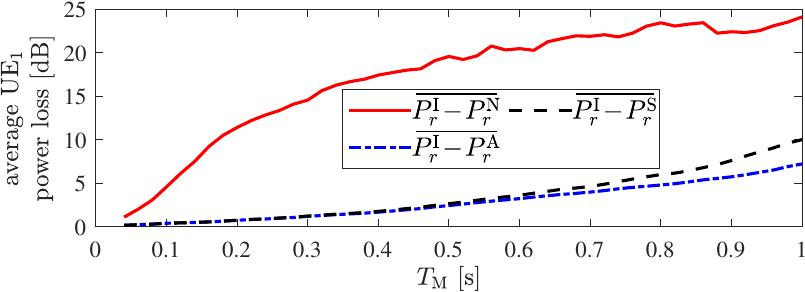}
    \\ (a) \vspace{.25cm} \\	\includegraphics[width=.235\textwidth]{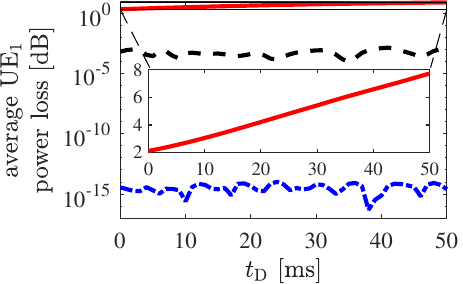}    \includegraphics[width=.245\textwidth]{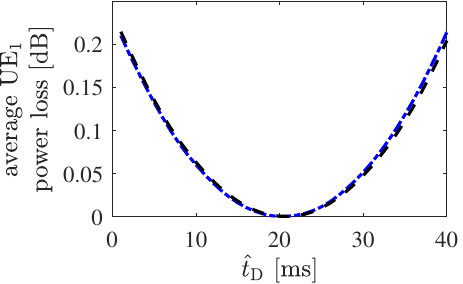}
    \\ \hspace{.05\textwidth} (b) \hspace{.2\textwidth} (c) \vspace{- .25cm} \\
    \caption{Average UE$_1$ power loss of the three control schemes w.r.t. 
    the ideal control ($P_r^{\textsc{i}}$), all as a function of the locator and observer sample rate in (a), as a function of $t_{\textsc{d}}$ delay in (b), and as a function of the estimated delay ${\hat t}_{\textsc{d}}$ given an exact delay of $t_{\textsc{d}}=20$~ms in (c). }
    \label{fig:SR}
\end{figure}

Fig.~\ref{fig:SR}a shows the performance of the approaches as a function of the sampling period $T_{\textsc{m}}$ of both the localizer and the observer, in the absence of measurement noise ($\sigma_n=0$). 
In all cases, increasing the sampling period leads to higher losses; however, the observer-plus-predictor approaches exhibit significantly smaller losses. 
Moreover, the OPA approach achieves the best performance, particularly for $T_{\textsc{m}}>0.5$~s. Fig.~\ref{fig:SR} also demonstrates that as $T_{\textsc{m}}$ approaches $T=1$~ms, all approaches converge toward the ideal scheme, with losses tending to zero. 
This condition, however, requires very high sampling rates from the localizer as well as increased communication load in the RIS control channel. 
Thus, selecting the sampling rate involves a tradeoff between minimizing power loss and limiting overhead in the RIS control channel, subject to localizer rate constraints.

\subsubsection{Different delays of RIS control}
Even in the absence of noise in the position measurements, power losses were evaluated as a function of the accumulated delay $t_{\textsc{d}}$ in Fig.~\ref{fig:SR}b. 
The naive approach exhibits a smooth monotonic increase in loss, starting at 2.1~dB for $t_{\textsc{d}}=0$ and reaching 7.7~dB at $t_{\textsc{d}}=50$~ms. 
In contrast, the proposed approaches incur negligible losses, with average values of $5.6\times10^{-4}$~dB and $3.9\times10^{-15}$~dB for OPS and OPA, respectively. 
This latter result highlights the accuracy of the OPA approach when measurement noise is absent.
\subsubsection{Effect of error in estimating accumulated delay}

Considering an exact value of $t_{\textsc{d}}=20$~ms, Fig.~\ref{fig:SR}c shows the average power loss as a function of the estimation error in the accumulated delay ${\hat t}_{\textsc{d}}$. 
The OPS and OPA approaches exhibit nearly identical and symmetrical behaviors. 
For estimation errors $|{\hat t}_{\textsc{d}}-t_{\textsc{d}}|\leq 5$~ms, the resulting losses are very small, below 0.02~dB. 
Even for absolute errors up to 20~ms, the power loss remains relatively low, reaching only 0.25~dB. 
Therefore, the proposed OPS and OPA approaches do not experience significant degradation due to errors in ${\hat t}_{\textsc{d}}$ when compared with typical losses in $P_r^{\textsc{i}}$.

\subsection{Numerical Results: Real Car Acceleration Scenario} \label{subse:ad_ae}

\begin{figure}[t]
    \centering
    \footnotesize \includegraphics[width=.49\textwidth]{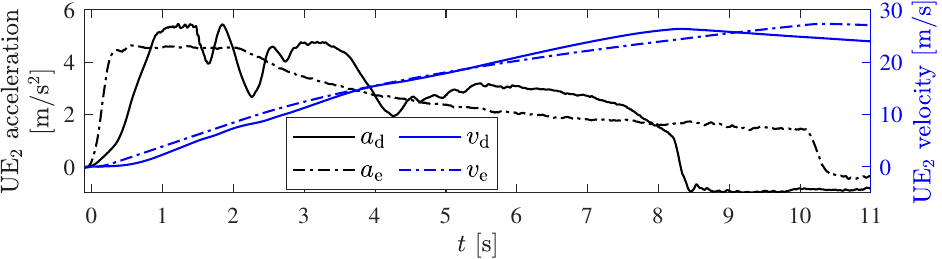}
    \\ (a) \vspace{.2cm} \\
    \includegraphics[width=.48\textwidth]{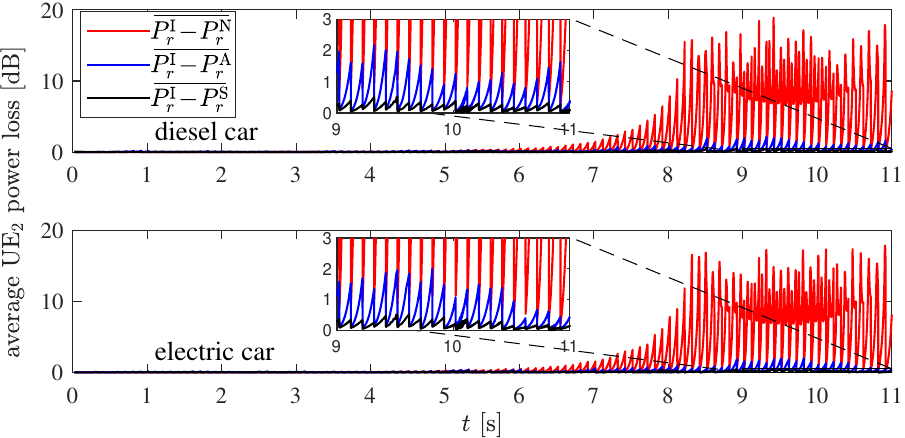}
    \\ (b) \vspace{-.25cm} \\
    \caption{Accelerations and speed from a 0 to 100~km/h test of a car with a diesel engine and another with an electric motor \cite{Meyn2015} in (a) and consequent average UE$_2$ power losses over time in (b).
    }
    \label{fig:golf}
\end{figure}

To evaluate more realistic motion scenarios, the accelerations measured from a car with a combustion engine and another with an electric engine were applied as input vectors to the kinematic model in \eqref{eq:ss_linear}. 
Fig.~\ref{fig:golf}a shows the acceleration and speed profiles for a 0--100~km/h test from \cite{Meyn2015}. 
These acceleration values are applied only along the $y$-axis, \textit{i.e.}, ${\bf u} = [0 \,\,-\!a_{\rm d}\,\,0]^{\textsc{t}}$~m/s$^2$ and ${\bf u} = [0 \,\,-\!a_{\rm e}\,\,0]^{\textsc{t}}$~m/s$^2$ for the diesel and electric engine cases, respectively.  

Assuming the cars start from rest ($v_{\rm d}=v_{\rm e}=0$) at position $[30\,\,150\,\,-\!30]^{\textsc{t}}$~m and with ${\sigma}_n=0.1$~m, the average power losses over time for ten realizations are shown in Fig.~\ref{fig:golf}b. 
The results indicate that the power losses for UE$_2$ in both the combustion and electric car cases are similar, reaching peaks greater than 18.9~dB under the naive control approach. 
In contrast, the proposed approaches achieve superior performance, particularly in the final stage of the movement when UE$_2$ is closer to the RIS and traveling at higher speeds.  
In this context, Table~\ref{tab:golf} highlights the effectiveness of the proposed approaches by reporting the average loss values during the final portion of the trajectory (from 7 to 11~s).  
This phase of the scenario involves faster variations in the RIS reflection angle, thereby imposing greater demands on the RIS controller.  

\begin{table}[h!]
\centering
\caption{Average losses in the final part of the car’s route (7 to 11~s)}
\vspace{-.2cm}
\label{tab:golf}
\begin{tabular}{|c|c|c|c|}
\hline
Engine type & $\overline{P_r^{\textsc{i}}-P_r^{\textsc{n}}}$ & $\overline{P_r^{\textsc{i}}-P_r^{\textsc{a}}}$ & $\overline{P_r^{\textsc{i}}-P_r^{\textsc{s}}}$ \\
\hline
diesel   & 6.32 & 0.35 & 0.14 \\
electric & 7.39 & 0.60 & 0.15 \\        
\hline               
\end{tabular}
\end{table}

\subsection{Numerical Results: Drone in Aggressive Piloting Scenario}
\begin{figure}[t]
    \centering
    \footnotesize
    \includegraphics[width=.245\textwidth]{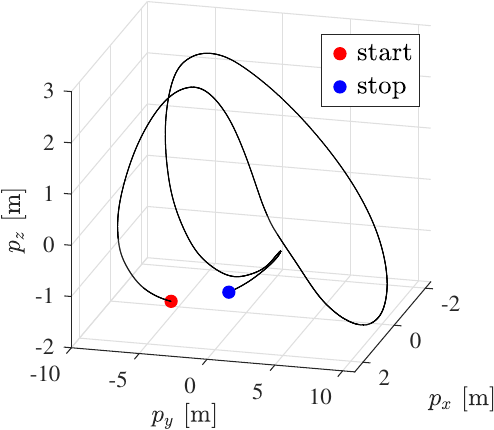}
    \includegraphics[width=.235\textwidth]{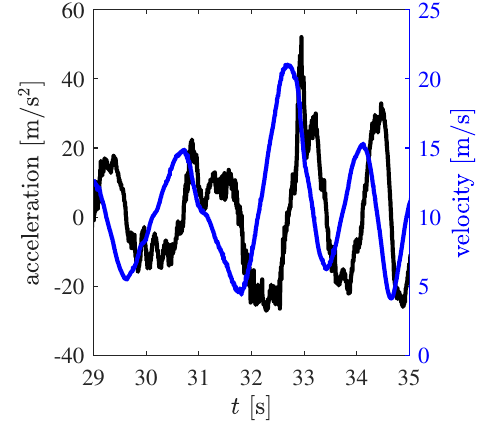}
    \\ (a) \hspace{.15\textwidth} (b)\vspace{.15cm} \\
    \includegraphics[width=.44\textwidth]{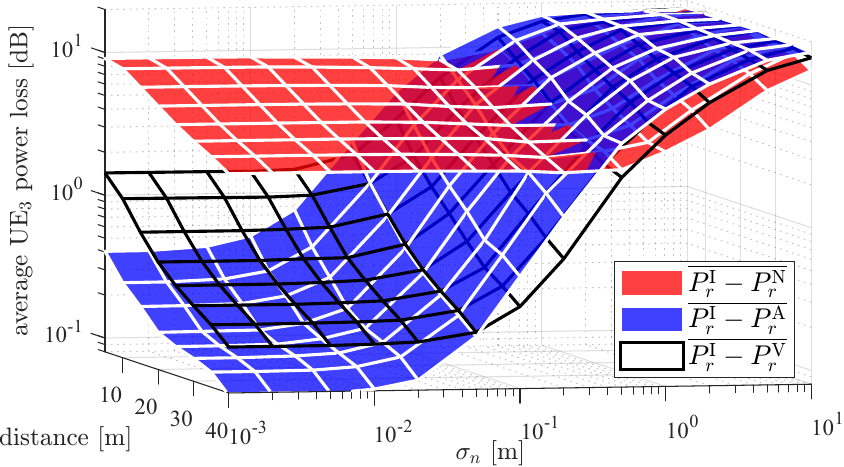}
    \\ (c) \vspace{-.25cm} \\
    \caption{Trajectory in (a) and acceleration and speed in (b) of a drone under aggressive piloting. 
    In (c), the average power loss in UE$_3$ as a function of the standard deviation of the localizer error $\sigma_n$ and the distance $x$ between the RIS and the center of the drone trajectory (Fig. \ref{fig:ris_scenario}). 
    }
    \label{fig:drone}
\end{figure}

To evaluate the proposed approach under extreme conditions, a dataset from an aggressively piloted racing drone was used as the position of UE$_3$. 
Specifically, the time interval between 29 and 35 seconds of flight dataset number 18a from \cite{Bosello2024} was selected, as it presents the highest speed and position gradients, with velocities exceeding 21~m/s. 
The dataset is recorded at 500~Hz. To enable direct comparison with the previous results, resampling was performed using spline interpolation to 1~kHz, thereby maintaining $T = 1$~ms. Since the longest displacement in the dataset occurs along the $x$-axis, for simulation convenience the $x$ and $y$ coordinates were swapped. 
This ensured a larger variation in the $y$-axis movement, making the scenario more challenging and requiring larger variations in the RIS reflection angles. 
Fig.~\ref{fig:drone}a shows the resulting trajectory, while Fig.~\ref{fig:drone}b shows the magnitudes of acceleration and speed. 
Notably, there is an instant when the acceleration exceeds 52~m/s$^2$, about 5.3 times greater than the acceleration due to gravity. 

The average power loss for UE$_3$ was then evaluated as a function of the standard deviation of the localizer measurements and for distances of 5, 10, ..., 40~m along the $x$-axis between the center of the drone trajectory and the RIS. 
To ensure line-of-sight (LoS) blocking between the BS and UE, the trajectory center was also shifted 10~m below the $z$-axis. 
Fig.~\ref{fig:drone}c presents the average of 10 realizations for each combination of $\sigma_n$ and distance $x$. 
The surfaces in the figure confirm the superior performance of the proposed approach, showing lower losses and greater tolerance to localizer errors as the trajectory center moves farther from the RIS. 
The observed behavior is consistent with that in Fig.~\ref{fig:PLs}, which considered uniformly accelerated motion.

\subsection{Numerical Results: Gauss-Markov Motion Scenario} 

The last scenario UE$_4$ uses the Gauss-Markov stochastic motion that combines directional memory and controlled randomness \cite{Yang2025}.
For this purpose, the UE movement block in Fig.~\ref{fig:ris_proposed_solution} is replaced, with its output given by the position:
\begin{equation}
    {{\bf{y}}}[k] = {{\bf{y}}}[k-1] + {{\bf{v}}}[k] T,
\end{equation}
with a speed vector ${{\bf{v}}}[k] =[\begin{matrix} v_{x,k} & v_{y,k} & v_{z,k} \end{matrix}]^{\textsc{t}}$ given by:
\begin{equation}
    {{\bf{v}}}[k] = \alpha {{\bf{v}}}[k-1] + (1-\alpha){\overline{\bf{v}}} + \sqrt{1-\alpha^2}{\bm \beta}[k],
\end{equation}
where ${\overline{\bf{v}}}=[\begin{matrix} \overline{v}_x & \overline{v}_y & \overline{v}_z \end{matrix}]^{\textsc{t}}$ is a vector with average speeds, ${\bm \beta}[k]=[\begin{matrix} \beta_{x,k} & \beta_{y,k} & \beta_{z,k} \end{matrix}]^{\textsc{t}}\sim \mathcal{N}(0,{\bf R}_{\beta})$ is a random signal that defines the unpredictability of the motion, with diagonal covariance matrix ${\bf R_{\beta}}= \text{diag}(\sigma_{\beta}^2,\,\sigma_{\beta}^2,\, \sigma_{\beta}^2)$ and standard deviation $\sigma_{\beta}$, finally, $\alpha \in [0,\,1]$ defines the degree of memory.

\begin{figure}[t]
    \centering
    \footnotesize
    \includegraphics[width=.44\textwidth]{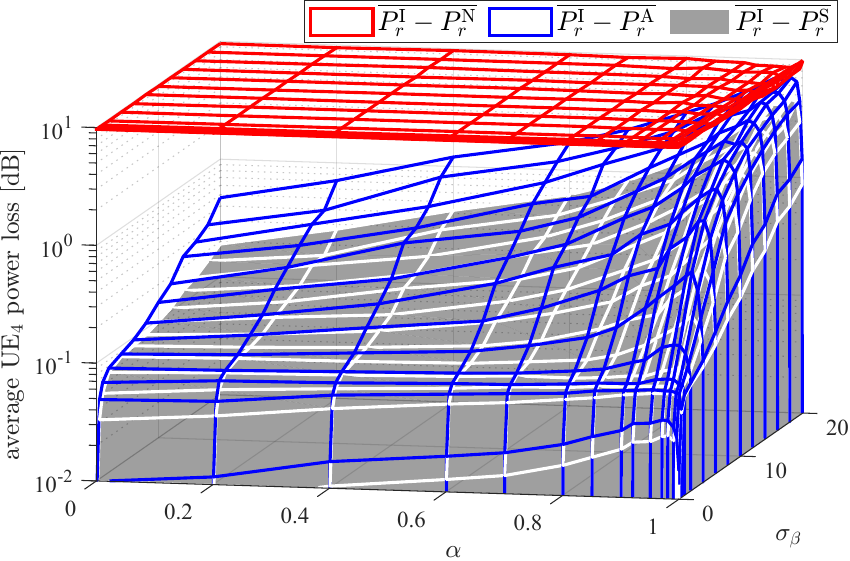}
    \vspace{-.25cm}
    \caption{Power losses for the Gauss-Markov motion scenario in UE$_4$ as a function of the memory factor $\alpha$ and the randomness factor $\sigma_\beta$.}
    \label{fig:gm}
\end{figure}

Fig.~\ref{fig:gm} presents the average power losses over 50 realizations, considering $0 \leq \alpha \leq 1$, $0.1 \leq \sigma_\beta \leq 20$~m, initial and average speeds ${\bf v}[k] = \overline{\bf v} = [20 \,-\!\!20 \, 0]^{\textsc{t}}$~m/s, and an initial position vector ${\bf y}[0] = [10 \,50 \, -\!\!30]^{\textsc{t}}$~m. 
The naive scheme exhibits consistently higher losses, with an approximately constant level near 10~dB, regardless of the values of $\alpha$ and $\sigma_\beta$. The results also show degradation in the performance of the OPA and OPS approaches as $\sigma_\beta$ increases, \textit{i.e.}, as the randomness of the movement grows. 
In this case, OPS again outperforms OPA. 
For both approaches, the losses increase with $\alpha$ (reduced memory) until approaching 1, where the largest losses occur. 
At this point, the movement becomes highly erratic due to the absence of a consistent trajectory pattern, representing the worst-case scenario for the proposed solutions. 
Beyond this point, the losses drop sharply as $\alpha$ approaches 1, since $\sqrt{1-\alpha^2}$ tends to zero, effectively eliminating the randomness of the movement.

\section{Conclusion}\label{sec:conclusion}

We have presented a new beamforming framework for RIS-assisted wireless communications based on a predictive, physics-based approach. 
The framework first employs a physics-based observer to estimate velocity and/or acceleration from sparse position updates provided by a low-rate secondary locator, such as a GNSS system. 
Subsequently, a physics-based predictor anticipates the UE's future position at a high rate, thereby compensating for the inherent RIS setup delay. Numerical simulations demonstrated significant reductions in power loss compared to naive control, enabled by delay mitigation and proactive adjustment of the RIS element phase shifts. Robustness was confirmed in typical scenarios, such as combustion and electric vehicles, as well as in more demanding conditions, including aggressive drone maneuvers and highly erratic trajectories. 
In noisy environments, the simpler OPS version outperformed the OPA approach, as it does not rely on acceleration estimation, a parameter more sensitive to secondary locator noise. Finally, the proposed approach demonstrates effectiveness, scalability, and low computational demand. 
Future work may further enhance performance by incorporating advanced estimation techniques, such as Kalman filtering or particle filtering, to mitigate the impact of measurement noise.

\appendices

\section{BS-RIS-UE backscattering
block}
\label{ap:ackscattering_block}

The power captured in the receiver antenna of the UE can be determined by:
\begin{equation}
\begin{array}{*{20}{l}}
P_r = \frac{P_t}{(4\pi)^2}\lambda^2 \sum\limits_{m = 1}^M {\sum\limits_{n = 1}^N { \left\{ G_t(\theta^{tx}_{m,n}) G_r(\theta^{rx}_{m,n}) \sigma (\theta^{t}_{m,n},\theta^{r}_{m,n}) \right.}}
\\
\qquad \qquad \qquad \times  \frac{|\Gamma_{m,n}|^2e^{2j \left( \angle \Gamma_{m,n} - k_0(|\bm{r}^t_{m,n}|+|\bm{r}^r_{m,n}|) \right)  }  }{|\bm{r}^t_{m,n}|^2|\bm{r}^r_{m,n}|^2} \}, 
\end{array}
\label{eq:Pr}
\end{equation}
where as shown in Fig. \ref{fig:ris_model},
$\theta^{r}_{m,n}\in [-\frac{\pi}{2},\,\frac{\pi}{2}]$ is the angle between the reflection vector in the $(n,m)$-th cell and $\bm{r}^{r}_{m,n}$, $0 \leq |\theta^{{tx}}_{m,n}| \leq \frac{\pi}{2}$ is the angle between ${\bm r}^{{tx}}$ and ${\bm r}^{t}_{m,n}$, and $0 < |\theta^{{rx}}_{m,n}| \leq \pi$ is the angle between ${\bm r}^{{rx}}$ and the $z$-axis.
$G_{t}(\theta_{m,n}^{{tx}})$ and $G_{r}(\theta_{m,n}^{{tx}})$ are the gains of the transmitting and receiving antennas,
and $\sigma(\theta^{r}_{m,n},\theta^{t}_{m,n})$ denotes the radar cross-section for transverse magnetic mode determined by \cite{Costa2021b}:
\begin{equation}
\resizebox{\linewidth}{!}{$
\sigma(\theta^{r}_{m,n},\theta^{t}_{m,n}) \!=\!
    \frac{4\pi D^4}{\lambda^2} \!\cos^2(\theta^{t}_{m,n}) {\rm sinc}^2\!\left( \frac{kD}{2}(\sin\theta^{r}_{m,n} \!\!-\! \sin\theta^{t}_{m,n}) \right).
    \label{eq:sigma}$}
\end{equation}

Considering the downlink case where the BS and RIS maintain fixed positions, and the RIS UPA has a square dimension, a symmetrical antenna with high directivity at the BS aligned to the center of the RIS is suitable. 
Its radiation pattern can be approximated by \cite{Balanis2016book}:
\begin{equation}
G_{t}(\theta_{m,n}^{{tx}}) \!=\! 
\frac{2\pi\cos^{q}(\theta_{m,n}^{{tx}})}{ \int_0^{2\pi}\int_0^{\pi/2 } { {\cos^{q}(\theta^{{tx}}) d\theta^{{tx}}} d\phi } } \!=\!
\frac{\cos^{q}(\theta_{m,n}^{{tx}})}{ \int_0^{\pi/2 } { {\cos^{q}(\theta^{{tx}}) d\theta^{{tx}}} } }, 
\label{eq:rad_pat_directive}
\end{equation}
where the higher the value of $q$, the higher the antenna's directivity.

At the receiver, if the UE is mobile and has variable orientations, a suitable radiation pattern is the omnidirectional type, which for the case of the half-wave dipole perpendicular to the $x$-$y$ plane, can be approximated by \cite{Balanis2016book}:
\begin{equation}
G_{r}(\theta_{m,n}^{{rx}}) = 
0.71199\frac{\cos^2\left( \frac{\pi}{2}\cos(\theta^{{rx}}_{m,n})\right)}{\sin^2(\theta^{{rx}}_{m,n})}.
\label{eq:rad_pat_omnidirectiona}
\end{equation}


\begin{thebibliography}{10}
\providecommand{\url}[1]{#1}
\csname url@samestyle\endcsname
\providecommand{\newblock}{\relax}
\providecommand{\bibinfo}[2]{#2}
\providecommand{\BIBentrySTDinterwordspacing}{\spaceskip=0pt\relax}
\providecommand{\BIBentryALTinterwordstretchfactor}{4}
\providecommand{\BIBentryALTinterwordspacing}{\spaceskip=\fontdimen2\font plus
\BIBentryALTinterwordstretchfactor\fontdimen3\font minus
  \fontdimen4\font\relax}
\providecommand{\BIBforeignlanguage}[2]{{%
\expandafter\ifx\csname l@#1\endcsname\relax
\typeout{** WARNING: IEEEtran.bst: No hyphenation pattern has been}%
\typeout{** loaded for the language `#1'. Using the pattern for}%
\typeout{** the default language instead.}%
\else
\language=\csname l@#1\endcsname
\fi
#2}}
\providecommand{\BIBdecl}{\relax}
\BIBdecl

\bibitem{Hong2023}
\BIBentryALTinterwordspacing
I.-P. Hong, ``Reviews based on the reconfigurable intelligent surface technical
  issues,'' \emph{Electronics}, vol.~12, no.~21, 2023. [Online]. Available:
  \url{https://www.mdpi.com/2079-9292/12/21/4489}
\BIBentrySTDinterwordspacing

\bibitem{He2022}
J.~He, H.~Wymeersch, M.~Di~Renzo, and M.~Juntti, ``Learning to estimate
  {RIS}-aided {mmWave} channels,'' \emph{IEEE Wireless Communications Letters},
  vol.~11, no.~4, pp. 841--845, 2022.

\bibitem{Saggese2023}
F.~Saggese, F.~Chiariotti, K.~Kansanen, and P.~Popovski, ``Efficient {URLLC}
  with a reconfigurable intelligent surface and imperfect device tracking,'' in
  \emph{ICC 2023 - IEEE International Conference on Communications}, 2023, pp.
  2722--2728.

\bibitem{Souza2024}
J.~H.~I. de~Souza, V.~Croisfelt, R.~Kotaba, T.~Abrão, and P.~Popovski,
  ``Uplink multiplexing of {eMBB/URLLC} services assisted by reconfigurable
  intelligent surfaces,'' \emph{IEEE Communications Letters}, vol.~28, no.~9,
  pp. 2206--2210, 2024.

\bibitem{Saggese2024}
F.~Saggese, V.~Croisfelt, R.~Kotaba, K.~Stylianopoulos, G.~C. Alexandropoulos,
  and P.~Popovski, ``On the impact of control signaling in {RIS}-empowered
  wireless communications,'' \emph{IEEE Open Journal of the Communications
  Society}, vol.~5, pp. 4383--4399, 2024.

\bibitem{Panichpapiboon2009}
S.~Panichpapiboon and W.~Pattara-atikom, ``An analysis of {GPS} sampling rates
  required in travel time estimation,'' in \emph{2009 IEEE Vehicular Networking
  Conference (VNC)}, 2009, pp. 1--6.

\bibitem{Gao2024}
L.~Gao, Z.~Fan, T.~He, J.~Lv, and X.~Zhang, ``A loosely coupled {INS/BDS}
  integrated navigation system,'' in \emph{2024 5th International Conference on
  Computer Vision, Image and Deep Learning (CVIDL)}, 2024, pp. 1054--1058.

\bibitem{teseo2025}
\BIBentryALTinterwordspacing
STMicroelectronics, ``{Teseo-ELE6A} – automotive quad-band {GNSS} module with
  open {SDK},'' Datasheet, Jun. 2025, {DS}14923 — Rev\,1. [Online].
  Available: \url{https://www.st.com/resource/en/datasheet/teseo-ele6a.pdf}
\BIBentrySTDinterwordspacing

\bibitem{zheng2023jrcup}
P.~Zheng, H.~Chen, T.~Ballal, M.~Valkama, H.~Wymeersch, and T.~Y. Al-Naffouri,
  ``{JrCUP}: Joint {RIS} calibration and user positioning for 6{G} wireless
  systems,'' \emph{IEEE Transactions on Wireless Communications}, vol.~23,
  no.~6, pp. 6683--6698, 2023.

\bibitem{zhang2023active}
Z.~Zhang, T.~Jiang, and W.~Yu, ``Active sensing for localization with
  reconfigurable intelligent surface,'' in \emph{ICC 2023-IEEE International
  Conference on Communications}.\hskip 1em plus 0.5em minus 0.4em\relax IEEE,
  2023, pp. 4261--4266.

\bibitem{teng2022bayesian}
B.~Teng, X.~Yuan, R.~Wang, and S.~Jin, ``Bayesian user localization and
  tracking for reconfigurable intelligent surface aided {MIMO} systems,''
  \emph{IEEE Journal of Selected Topics in Signal Processing}, vol.~16, no.~5,
  pp. 1040--1054, 2022.

\bibitem{Hu2025}
D.~Hu, J.~Nakazato, K.~Maruta, R.~Dinis, O.~A. Aghda, and M.~Tsukada, ``{OTFS}
  based {RIS}-assisted vehicle positioning and tracking in {V2X} scenario,''
  \emph{IEEE Transactions on Communications}, pp. 1--1, 2025.

\bibitem{wang2023b}
J.~Wang, W.~Tang, S.~Jin, C.-K. Wen, X.~Li, and X.~Hou, ``Hierarchical
  codebook-based beam training for {RIS}-assisted {mmWave} communication
  systems,'' \emph{IEEE Transactions on Communications}, vol.~71, no.~6, pp.
  3650--3662, 2023.

\bibitem{wang2023uav}
J.~Wang, X.~Wang, X.~Liu, C.-T. Cheng, F.~Xiao, and D.~Liang, ``Trajectory
  planning of {UAV}-enabled data uploading for large-scale dynamic networks: A
  trend prediction based learning approach,'' \emph{IEEE Transactions on
  Vehicular Technology}, vol.~72, no.~6, pp. 8272--8277, 2023.

\bibitem{hu2024location}
X.~Hu, Y.~Tian, Y.~H. Kho, B.~Xiao, Q.~Li, Z.~Yang, Z.~Li, and W.~Li,
  ``Location prediction using {B}ayesian optimization {LSTM} for {RIS}-assisted
  wireless communications,'' \emph{IEEE Transactions on Vehicular Technology},
  vol.~73, no.~10, pp. 15\,156--15\,171, 2024.

\bibitem{akrout2023domain}
M.~Akrout, A.~Feriani, F.~Bellili, A.~Mezghani, and E.~Hossain, ``Domain
  generalization in machine learning models for wireless communications:
  Concepts, state-of-the-art, and open issues,'' \emph{IEEE Communications
  Surveys \& Tutorials}, vol.~25, no.~4, pp. 3014--3037, 2023.

\bibitem{khalil2015nonlinear}
H.~K. Khalil, \emph{Nonlinear Systems}, 3rd~ed.\hskip 1em plus 0.5em minus
  0.4em\relax Prentice Hall, 2015.

\bibitem{Costa2021a}
F.~Costa and M.~Borgese, ``Circuit modelling of reflecting intelligent
  surfaces,'' in \emph{2021 IEEE 22nd International Workshop on Signal
  Processing Advances in Wireless Communications (SPAWC)}, 2021, pp. 546--550.

\bibitem{Costa2021b}
------, ``Electromagnetic model of reflective intelligent surfaces,''
  \emph{IEEE Open Journal of the Communications Society}, vol.~2, pp.
  1577--1589, 2021.

\bibitem{skyworks2015}
\emph{Varactor {SPICE} Models for {RF VCO} Applications}, Skyworks Solutions
  Inc, Invine, CA, august 2015, application note.

\bibitem{Liu2021b}
Y.~Liu, X.~Liu, X.~Mu, T.~Hou, J.~Xu, M.~Di~Renzo, and N.~Al-Dhahir,
  ``Reconfigurable intelligent surfaces: Principles and opportunities,''
  \emph{IEEE Communications Surveys \& Tutorials}, vol.~23, no.~3, pp.
  1546--1577, 2021.

\bibitem{Becker2023book}
A.~Becker, \emph{Kalman filter from the ground up}.\hskip 1em plus 0.5em minus
  0.4em\relax KilmanFilter. NET, 2023.

\bibitem{Wu1990}
L.~E.~E. Jiangjiexing~Wu, Liang~Fan, ``Three-point backward finite-difference
  method for solving a system of mixed hyperbolic—parabolic partial
  differential equations,'' \emph{Computers \& chemical engineering}, vol.~14,
  no.~6, pp. 679--685, 1990.

\bibitem{Carneiro2016}
J.~Falcão~Carneiro and F.~Gomes De~Almeida, ``On the influence of velocity and
  acceleration estimators on a servopneumatic system behaviour,'' \emph{IEEE
  Access}, vol.~4, pp. 6541--6553, 2016.

\bibitem{Meyn2015}
J.-P. Meyn, ``The kinematic advantage of electric cars,'' \emph{European
  Journal of Physics}, vol.~36, no.~6, p. 065037, 2015.

\bibitem{An2023}
X.~An, R.~Ziebold, and C.~Lass, ``From {RTK} to {PPP-RTK}: Towards real-time
  kinematic precise point positioning to support autonomous driving of inland
  waterway vessels,'' \emph{GPS Solutions}, vol.~27, no.~2, p.~86, 2023.

\bibitem{Bosello2024}
\BIBentryALTinterwordspacing
M.~Bosello, D.~Aguiari, Y.~Keuter, E.~Pallotta, S.~Kiade, G.~Caminati,
  F.~Pinzarrone, J.~Halepota, J.~Panerati, and G.~Pau, ``Race against the
  machine: A fully-annotated, open-design dataset of autonomous and piloted
  high-speed flight,'' \emph{IEEE Robotics and Automation Letters}, vol.~9,
  no.~4, pp. 3799--3806, 2024. [Online]. Available:
  \url{https://github.com/tii-racing/drone-racing-dataset}
\BIBentrySTDinterwordspacing

\bibitem{Yang2025}
S.~Yang, W.~Zhao, C.-M. Wang, W.-Y. Dong, and X.~Ju, ``Betweenness centrality
  based dynamic source routing for flying ad hoc networks in marching
  formation,'' \emph{IEEE Transactions on Vehicular Technology}, pp. 1--8,
  2025.

\bibitem{Balanis2016book}
C.~A. Balanis, \emph{Antenna theory: analysis and design}, 4th~ed.\hskip 1em
  plus 0.5em minus 0.4em\relax John Wiley \& sons, 2016.

\end{thebibliography}
\end{document}